\documentclass[conference]{IEEEtran}
\IEEEoverridecommandlockouts

\usepackage[utf8]{inputenc}
\usepackage{amsmath}
\usepackage{amssymb}
\usepackage{amsfonts}
\usepackage{amsthm}
\usepackage{optidef}
\usepackage{multirow}
\usepackage{threeparttable}
\usepackage{todonotes}
\usepackage{array}
\usepackage{algorithm}
\usepackage{algorithmic}
\usepackage{caption} % multi figures
\usepackage{subcaption} % multi figures
\usepackage{cite}
\usepackage{verbatim} % comment
\usepackage{setspace} % line space
\usepackage[bookmarks=false]{hyperref}
\hypersetup{colorlinks=true,linkcolor=blue,citecolor=blue}
\allowdisplaybreaks[4] % equation in two pages

\title{Fault-Aware Design for Reconfigurable Holographic Surface-Aided ISAC Systems}
\date{}

\author{
\IEEEauthorblockN{
Lu Wang\IEEEauthorrefmark{1},
Mohamadreza Delbari\IEEEauthorrefmark{1},
Gui Zhou\IEEEauthorrefmark{2},
Luis F. Abanto-Leon\IEEEauthorrefmark{3},
Matthias Hollick\IEEEauthorrefmark{1},
and Vahid Jamali\IEEEauthorrefmark{1}
}
\IEEEauthorblockA{\IEEEauthorrefmark{1}Technical University of Darmstadt, Darmstadt, Germany, \IEEEauthorrefmark{2}Huazhong University of Science and Technology, Wuhan, China
}
\IEEEauthorblockA{\IEEEauthorrefmark{3}Ruhr University Bochum, Bochum, Germany
}
\thanks{This work was supported by the Deutsche Forschungsgemeinschaft (DFG) mmCell project under Grant 416765679, mmV2X project under Grant 453080125, and HyRIS project under Grant 455077022. The work of Delbari and Jamali was supported in part by the LOEWE initiative (Hesse, Germany) within the emergenCITY Centre under Grant LOEWE/1/12/519/03/05.001(0016)/72, and in part by the German Federal Ministry for Research, Technology and Space (BMFTR) under the program of “Souverän. Digital. Vernetzt.” joint project Open6GHub plus (Project-ID 16KIS2407).}
}

\begin{document}
\maketitle

\begin{abstract}

Reconfigurable holographic surface (RHS)-aided integrated sensing and communication (ISAC) systems hold great promise for achieving both sensing and communication with low hardware costs and high energy efficiency. However, existing works largely overlook practical hardware impairments in RHSs, particularly faulty RHS elements with uncontrollable amplitudes, which degrade system performance if left unaddressed. This work aims to fill the gap by \textit{i) quantifying the impact of faulty RHS elements on ISAC performance and ii) optimizing the functional RHS elements to preserve the ISAC performance}. Specifically, we derive the misspecified Cramér-Rao bound (MCRB) for sensing and the signal-to-interference-and-noise ratio (SINR) for communication to measure the performance loss caused by faulty elements. We then formulate an optimization problem that minimizes MCRB, subject to constraints on SINR, transmit power budget, and RHS amplitude. The high non-convexity of the formulated problem poses a significant challenge, which we address by reformulating and proposing a block coordinate descent-based solution incorporating majorization-minimization and successive convex approximation techniques. Simulation results verify that the proposed approach achieves an average 13.7\% performance gain compared to the fault-unaware benchmark.

\end{abstract}

%\begin{IEEEkeywords}
%ISAC, RHS, faulty elements, MCRB, SINR
%\end{IEEEkeywords}
\section{Introduction}
\label{Introduction}
Integrated sensing and communication (ISAC) has emerged as a key paradigm for future 6G networks, enabling the unified design of sensing and communication functionalities within a single system. This integration is expected to support various emerging applications, such as autonomous driving and smart homes~\cite{bSHLu,bFanoverview}. To simultaneously achieve high communication throughput and accurate sensing performance, ISAC systems are envisioned to employ massive or ultra-massive multiple-input multiple-output (MIMO) architectures, typically implemented using phased-array structures~\cite{bRHSOver1RqD}. While increasing the number of antennas can significantly enhance ISAC performance, it also leads to a substantial rise in power and energy consumption as well as hardware cost, due to a large-scale phase array consisting of power amplifiers and phase shifters~\cite{bRHSOver2ByDi}. Moreover, the large physical aperture associated with such arrays poses additional challenges for practical deployment.

To address these challenges, the reconfigurable holographic surface (RHS) technology has emerged, offering advantages such as reduced power consumption, lower hardware costs, and a smaller size, all while maintaining the same number of antennas/elements~\cite{bRHSOver1RqD,bRHSOver2ByDi,bRHSISACover}. This is because the RHS can be realized by simpler hardware, including positive-intrinsic-negative (PIN)-diodes and voltage controllers~\cite{bRHSOver2ByDi}. Furthermore, the inter-element spacing of RHSs can be smaller than a half wavelength, enabling a denser and more compact physical structure~\cite{bRHSISACtech1}. Owing to the aforementioned advantages, RHS-aided ISAC systems are promising to realize an energy-efficient, cost-effective, yet performance-guaranteed framework.

RHS elements, like all hardware devices, are susceptible to impairments due to various reasons, such as aging after long-term usage, natural catastrophes, etc.~\cite{bHWIall}. For surfaces with a large number of elements, directly replacing the entire hardware when partial elements become faulty is neither cost-effective nor energy-efficient. A more practical approach is therefore to exploit the remaining functional elements judiciously to mitigate the performance degradation caused by such faults. To date, numerous studies have investigated hardware impairments in phase array antennas and reconfigurable intelligent surface (RIS), including imperfections in antenna and reflecting elements. For instance, the diagnosis of faulty elements, including the location, induced attenuation, and phase shift change, was studied using compressed sensing for phase array antennas in~\cite{bPAfault1,bPAfault2} and for RIS in~\cite{bfaultyDetect1,bfaultyDetect2}. To achieve fast and robust fault diagnosis, the work~\cite{bPAfault3} proposed a deep neural network-based method for phase array antennas, which achieves over 80\% accuracy in only milliseconds.

On the other hand, fault-aware measures have been taken to ensure reliable system performance in the presence of antenna or element failure. The work~\cite{bPAfault4} compensated for the performance degradation due to the planar array failure by deriving the element excitations. In~\cite{bPAfault5}, the authors proposed an integrated framework for diagnosing and correcting faulty antenna arrays using a Bayesian compressed sensing method. In RIS scenarios, the work~\cite{bNairyRIS} mitigated the information leakage due to RIS failure by optimizing the signal-to-leakage-and-noise ratio of the communication user. The authors in~\cite{bLW} enhanced the sensing performance while guaranteeing the communication performance for the RIS-aided ISAC system in the presence of faulty RIS elements.

However, the aforementioned failure models cannot be directly applied to RHS-aided systems due to different working principles. There have been works studying RHS-aided systems with hardware impairments at the transceiver~\cite{bRHSHWI1,bRHSHWI2,bRHSHWI3}, but they did not account for impairments at the RHS. How to guarantee both sensing and communication performance in the presence of RHS failures remains important and unexplored. This work aims to fill this research gap, focusing on mitigating the negative impact of faulty RHS elements on the ISAC system. To realize this goal, we optimize the amplitudes of the functional RHS elements to improve both sensing and communication performances. Specifically, we derive the misspecified Cramér-Rao bound (MCRB) and signal-to-interference-and-noise ratio (SINR) to measure the sensing and communication performance degradation, respectively, due to RHS failures. Our contributions are summarized as follows: (i) This is the first work to mitigate the negative impact of faulty RHS elements on ISAC systems, measured by the derived MCRB and SINR. (ii) To reach this goal, we formulate an optimization problem minimizing the sensing MCRB while satisfying the communication SINR, transmit power, and RHS hardware constraints. By jointly designing digital beamforming and holographic beamforming, the formulated problem is solved using the proposed block coordinate descent (BCD) algorithm based on majorization-minimization (MM), successive convex approximation (SCA), and penalization techniques. (iii) Simulation results confirm that the proposed solution reduces the performance loss by 13.7\% on average compared to the naive scheme, which neglects the presence of faulty RHS elements.

%The rest of the paper is organized as follows. Section \uppercase\expandafter{\romannumeral2} introduces the system model, followed by the MCRB derivation in Section \uppercase\expandafter{\romannumeral3}. Sections \uppercase\expandafter{\romannumeral4} and \uppercase\expandafter{\romannumeral5} present the formulated problem and the proposed algorithm. Simulation results and analysis are provided in Section \uppercase\expandafter{\romannumeral6}, followed by the conclusion in Section \uppercase\expandafter{\romannumeral7}.

%\textit{Notation}: Bold uppercase and lowercase letters (e.g. ${\mathbf{X}}_1$ and ${\mathbf{x}}_1$) denote matrices and vectors, respectively. $\mathbb{E}\{ \cdot \}$ and ${\mathcal O}(\cdot)$ represent statistical expectation and computational complexity order. ${\rm{tr}}(\cdot)$, $(\cdot)^{-1}$, ${(\cdot)^{\rm{T}}}$ and ${(\cdot)^{\rm{H}}}$ denote the trace, inverse, transpose, and Hermitian, respectively. ${{\bf{X}}_1} \succeq 0$ and ${{\bf{X}}_1} \prec 0$ indicate positive semi-definite (PSD) and negative definite matrices, respectively. ${\| \cdot \|_{\rm{F}}}$, ${\rm{rank}}(\cdot)$, and ${\rm{diag}}(\cdot)$ denote the Frobenius norm, rank, and diagonal operator. ${\mathbb{C}}$ and ${\mathbb{R}}$ denote the complex and real fields, and ${\rm{Re}}(\cdot)$ denotes the real part.
\section{System Model}
\addtolength{\topmargin}{0.05in}

\begin{figure}[!t]
    \centering
    \vspace{0.01cm}
    \includegraphics[width=0.85\linewidth]{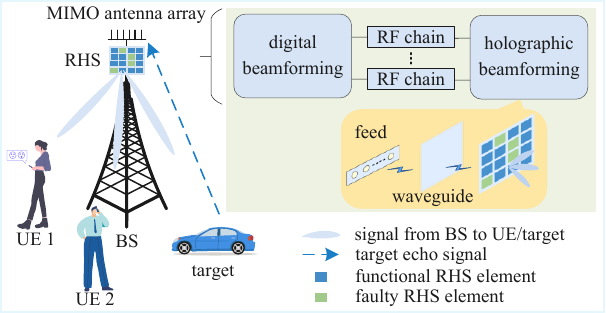}
    \caption{\small{RHS-aided ISAC system with faulty elements.}}
    \label{Fig_sysm_RHS}
    \vspace{-0.5cm}    
\end{figure}

We consider an RHS-aided ISAC system, where the base station (BS) simultaneously communicates with $K$ single-antenna user equipments (UEs) and senses one target, as shown in Fig.~\ref{Fig_sysm_RHS}. The uniform planar RHS, with a dense inter-element spacing \textit{less than a half wavelength}, consists of $N_{RF}$ feeds, a parallel-plate waveguide, and $N_t = N_{ty}N_{tz}$ radiating elements. The transmitted symbols are first processed by digital beamforming and radio frequency (RF) chains, and then fed to the RHS for amplitude-controlled holographic beamforming. We study a practical scenario where a subset of RHS elements is faulty. Our goal is to optimize the functional RHS elements to ensure the ISAC performance.

\subsection{Signal Model}
The baseband transmit signal at time slot $t$ is expressed as
\begin{equation}\label{Xbb}
{\bf{x}}\left( t \right) = \sum\nolimits_{k = 1}^K {{{\bf{w}}_k}{c_k}\left( t \right)}  + {\bf{s}}\left( t \right) = {\bf{Wc}}\left( t \right) + {\bf{s}}\left( t \right),
\end{equation}
where ${\bf{c}}(t) = {[ {{c_1}(t),...,{c_K}(t)} ]^{\rm{T}}}\in {\mathbb{C}^{K \times 1}}$ refers to the symbols transmitted to $K$ UEs, modeled as independently and identically distributed random variables, satisfying $\mathbb{E} \{ {{\bf{c}}(t){\bf{c}}{{(t)}^{\rm{H}}}} \} = {{\bf{I}}_K}$. ${\bf{s}}(t)$ is the dedicated signal introduced for sensing to achieve full degree of freedom, with its covariance matrix given by ${{\bf{R}}_s} = {\textstyle{1 \over T}}\sum\nolimits_{t = 1}^T {{\bf{s}}(t){\bf{s}}{{(t)}^{\rm{H}}}} \succeq {\bf{0}}$, for a large value of $T$. They satisfy $\mathbb{E} \{ {{\bf{c}}(t){\bf{s}}{{(t)}^{\rm{H}}}} \} = {{\bf{0}}_{K \times {N_{RF}}}}$. ${\bf{W}} = [ {{{\bf{w}}_1},...,{{\bf{w}}_K}}] \in {\mathbb{C}^{N_{RF} \times K}}$ denotes the digital beamforming matrix for $K$ UEs, with the $k$-th column ${{\bf{w}}_k}\in {\mathbb{C}^{N_{RF} \times 1}}$ corresponding to UE $k$'s beamforming vector. After ${\bf{x}}(t)$ being processed by digital beamforming and $N_{RF}$ RF chains, it is fed to the RHS for holographic beamforming. Specifically, let ${\boldsymbol{\psi}} = {[ {{\psi _1},...,{\psi _{{N_t}}}}]^{\rm{T}}} \in \mathbb{C}^{N_t \times 1}$ collect the amplitudes of all RHS elements, where ${\psi _n} \in [0,1]$ is the amplitude of element $n$. ${\boldsymbol {\Psi}}  = {\rm{diag}}({\boldsymbol{\psi}}^{\rm{T}})\in {\mathbb{C}^{N_t \times N_t}}$ denotes the amplitude matrix. Let ${\bf{Q}}\in {\mathbb{C}^{N_t \times N_{RF}}}$ denote the phase shift coefficients from feeds to RHS, with the $(n,l)$-th entry expressed as ${q_{n,l}} = {e^{ - j2\pi \kappa {d_{n,l}}/\lambda }}$, where $\kappa$ is the refractive index, $d_{n,l}$ is the distance from the $l$-th feed to $n$-th RHS element, and $\lambda$ is the wavelength of reference wave. The transmit signal after hybrid beamforming is given by~\cite{bRHSISACtech1}
\begin{equation}\label{Xhb}
{{\bf{x}}_1}\left( t \right) = {\bf{\Psi Qx}}\left( t \right) = {\bf{\Psi Q}}\left( {{\bf{Wc}}\left( t \right) + {\bf{s}}\left( t \right)} \right).
\end{equation}

\subsection{RHS Model}

\subsubsection{Perfect RHS Model}
\label{secPerfect}
When the presence of faulty elements is not detected, the perfect RHS model with all functional elements is assumed, leading to a mismatch in the RHS model. In this case, the amplitude of each element is denoted by ${{\tilde \psi }_n}$, satisfying ${{\tilde \psi } _n} \in [0,1], \forall n \in \{1,...,N_t\}$.

\subsubsection{Faulty RHS Model}
\label{secTrue}
In the presence of RHS failures, the amplitudes of RHS elements are denoted by ${{\bar \psi }_n}$. Among the total $N_t$ of RHS elements, $F$ are faulty, with the remaining $W = N_t-F$ being functional. The index sets of the faulty and functional elements are denoted by ${\mathcal S_F}$ and ${\mathcal S_W}$, respectively. The faulty elements are assumed to be randomly distributed among $N_t$ elements, with their locations detected and known by the methods proposed in~\cite{bfaultyDetect1,bfaultyDetect2}. The amplitudes of these faulty elements are unknown and modeled as independent random variables uniformly distributed over $[0,1]$. Note that although the amplitudes of both faulty and functional elements satisfy ${{\bar \psi }_n} \in [0,1]$, only the amplitudes of functional elements are design variables and can be optimized to improve system performance. In contrast, the amplitudes of faulty elements are uncontrollable, inducing random attenuation and degrading the quality of the generated beams.

\subsection{Communication Model}
\label{COMSINR}
For the $k$-th UE, the received signal from the BS at time slot $t$ is expressed as follows
\begin{equation}\label{Yck}
\begin{array}{*{20}{l}}
{{y_{c,k}}\left( t \right) = {\bf{h}}_k^{\rm{H}}{\bf{x}}_1\left( t \right) = \underbrace {{\bf{h}}_k^{\rm{H}} {\bf{\bar \Psi Q}}{{\bf{w}}_k}{c_k}\left( t \right)}_{{\rm{desired}}\;{\rm{signal}}} + }\\
{\underbrace {\sum\limits_{i = 1,i \ne k}^K {{\bf{h}}_k^{\rm{H}}{\bf{\bar  \Psi Q}}{{\bf{w}}_i}{c_i}\left( t \right)} }_{{\rm{inter - user}}\;{\rm{interference}}} + \underbrace {{\bf{h}}_k^{\rm{H}}{\bf{\bar \Psi Q}}{{\bf{s}}\left( t \right)}}_{{\rm{sensing}}\;{\rm{interference}}} + {n_k}(t),}
\end{array}
\end{equation}
where for UE $k$, ${{\bf{h}}_k}\in {\mathbb{C}^{N_t \times 1}}$ denotes the BS-UE channel, and ${n_k}(t)$ is complex Gaussian noise. Signals emitted through faulty RHS elements undergo unwanted amplitude distortions. The SINR is used to evaluate each UE's performance under RHS faults, aiming to enhance the desired signal power and suppress interference from other UEs and the target, given by
\begin{equation}\label{CSINR}
{\gamma _k} = \frac{{{{\left| {{\bf{h}}_k^{\rm{H}}{\bf{\bar \Psi Q}}{{\bf{w}}_k}} \right|}^2}}}{{\sum\limits_{i \ne k}^K {{{\left| {{\bf{h}}_k^{\rm{H}} {\bf{\bar \Psi Q}}{{\bf{w}}_i}} \right|}^2}} + {\bf{h}}_k^{\rm{H}} {\bf{\bar \Psi Q}}{{\bf{R}}_s}{{\bf{Q}}^{\rm{H}}}{{\bf{\bar \Psi}} ^{\rm{H}}}{\bf{h}}_k + \sigma _k^2}}.
\end{equation}

\subsection{Sensing Model}
%%%%%%%%%%%%%%% Channel %%%%%%%%%%%%%%%%%
For sensing, this work focuses on estimating the target angle of departure (AoD), denoted as ${\boldsymbol{\phi}} = {\left[ {{\phi _e},{\phi _a}} \right]^{\rm{T}}}$, where ${\phi _e}$ and ${\phi _a}$ are the elevation and azimuth AoD, respectively. The sensing target response matrix is expressed as
\begin{equation}\label{GTRM}
{\bf{G}} = \alpha {\bf{a}}\left( {{\phi _e},{\phi _a}} \right){{\bf{a}}^{\rm{H}}}\left( {{\phi _e},{\phi _a}} \right),
\end{equation}
where $\alpha$ is the channel coefficient. ${\bf{a}}\left( {{\phi _e},{\phi _a}} \right) = \left[ {{a}{{\left( {{\phi _e},{\phi _a}}  \right)}_1},...,{a}{{\left( {{\phi _e},{\phi _a}} \right)}_{N_t}}} \right]^{\rm{T}} \in {\mathbb{C}^{N_t \times 1}}$ denotes the steering vector with respect to (w.r.t) RHS deployed in the y-z plane.
Each element is expressed as
\begin{equation}\label{SVa_r}
{a}{\left( {{\phi _e},{\phi _a}} \right)_n} = {e^{-j\kappa_1 \left[ {{d_y}\cos \left( {{\phi _e}} \right)\sin \left( {{\phi _a}} \right){n_y} + {d_z}\sin \left( {{\phi _e}} \right){n_z}} \right]}},
\end{equation}
where $\kappa_1  = {{2\pi } \mathord{\left/ {\vphantom {{2\pi } \lambda }} \right. \kern-\nulldelimiterspace} \lambda }$ is the wave number. $d_y$ and $d_z$ are the spacing between adjacent RHS elements along y-axis and z-axis, which are indexed by ${n_y} \in \left\{ {0,1,...,{N_{ty}} - 1} \right\}$ and ${n_z} \in \left\{ {0,1,...,{N_{tz}} - 1} \right\}$, respectively.
To estimate the AoD, the received echo signals at the BS are analyzed over a coherent time block with $T$ slots, namely
\begin{equation}\label{YsT}
{{\bf{Y}}_s} = {\bf{G \bar \Psi QX}} + {{\bf{N}}_s},
\end{equation}
where ${\bf{X}} = \left[ {{\bf{x}} \left( 1 \right),...,{\bf{x}}\left( T \right)} \right] \in {\mathbb{C}^{N_t \times T}}$. ${{\bf{N}}_s} \in {\mathbb{C}^{N_t \times T}}$ is the noise with each entry obeying the complex Gaussian distribution with zero mean and variance $\sigma_s^2$.

%%%%%%%%%%%%%%%%%%%% MCRB %%%%%%%%%%%%%%%%%%%%%%%
\subsection{MCRB Derivation}
To measure the accuracy of AoD estimation, the Cramér-Rao bound (CRB) is commonly used to provide a lower bound on the mean square error of unbiased estimation. However, when faulty RHS elements exist but remain unknown, assuming a perfect RHS model leads to a RHS model mismatch. In this case, CRB becomes inappropriate as it does not account for the RHS model mismatch. To address this issue, we adopt the MCRB to explicitly capture the RHS model mismatch \cite{bMCRB} and analyze the degradation in AoD estimation performance.
 
\subsubsection{Pseudo-true Parameter ${{\boldsymbol{\eta }}_0}$}
Under the faulty RHS model, the unknown parameters are denoted by ${\boldsymbol{\bar \eta }} = {\left[ {{{\bar \phi }_e},{{\bar \phi }_a},{\rm{Re}}\left( {\bar \alpha } \right),{\rm{Im}}\left( {\bar \alpha } \right)} \right]^{\rm{T}}} \in {\mathbb{R}^{4\times 1}}$. Vectorizing (\ref{YsT}) yields
\begin{equation}\label{YsVector}
{{\bf{y}}_s} = {\rm{vec}}\left( {{{\bf{Y}}_s}} \right) = {\rm{vec}}\left( \bf{G \bar \Psi QX} \right) + {\rm{vec}}\left( {{{\bf{N}}_s}} \right) = {\boldsymbol{\bar \mu }} + {{\bf{n}}_s},
\end{equation}
where ${\boldsymbol{\bar \mu }} = {\rm{vec}}\left( \bf{G \bar \Psi QX} \right) \in {\mathbb{C}^{{N_t}T\times 1}}$. The true probability density function (PDF) of ${{\bf{y}}_s}$ under faulty RHS model is
\begin{equation}\label{PDFtrue}
p\left( {{{\bf{y}}_s}} \right) = \frac{1}{{\sqrt {{{\left( {2\pi \sigma _s^2} \right)}^{{N_t}T}}} }}\exp \left( { - \frac{{{{\left\| {{{\bf{y}}_s} - {\boldsymbol{\bar \mu }}} \right\|}^2}}}{{2\sigma _s^2}}} \right).
\end{equation}
Under the perfect RHS assumption, the parameter vector is ${\boldsymbol{\eta }} = {\left[ {{\phi _e},{\phi _a},{\rm{Re}}\left( \alpha  \right),{\rm{Im}}\left( \alpha  \right)} \right]^{\rm{T}}} \in {\mathbb{R}^{4\times 1}}$ with noise-free observation ${\boldsymbol{\tilde \mu }} = {\rm{vec}}( {\bf{G \tilde \Psi QX}}) \in {\mathbb{C}^{{N_t}T\times 1}}$ and misspecified PDF of the received echo
\begin{equation}\label{PDFmisspec}
\tilde p\left( {{{\bf{y}}_s}\left| {\boldsymbol{\eta }} \right.} \right) = \frac{1}{{\sqrt {{{\left( {2\pi \sigma _s^2} \right)}^{{N_t}T}}} }}\exp \left( { - \frac{{{{\left\| {{{\bf{y}}_s} - {\boldsymbol{\tilde \mu }}\left( {\boldsymbol{\eta }} \right)} \right\|}^2}}}{{2\sigma _s^2}}} \right).
\end{equation}
Then, we define the pseudo-true parameter ${{\boldsymbol{\eta }}_0}\in {\mathbb{R}^{4\times 1}}$ as the minimizer of the Kullback-Leibler (KL) divergence between ${p\left( {{{\bf{y}}_s}} \right)}$ and ${\tilde p\left( {{{\bf{y}}_s}\left| {\boldsymbol{\eta }} \right.} \right)}$, denoted by ${D_{KL}}$, namely
\begin{equation}\label{PseudoP}
{{\boldsymbol{\eta }}_0} = \mathop {\arg \min }\limits_{\boldsymbol{\eta }} {D_{KL}}\left( {p\left( {{{\bf{y}}_s}} \right)\left\| {\tilde p\left( {{{\bf{y}}_s}\left| {\boldsymbol{\eta }} \right.} \right)} \right.} \right).
\end{equation}
${{\boldsymbol{\eta }}_0}$ finds the PDF of the perfect RHS model that is nearest to the true PDF ${p\left( {{{\bf{y}}_s}} \right)}$ in the KL divergence sense\cite{bMCRBorigin}. ${{\boldsymbol{\eta }}_0}$ can be obtained by the method proposed in~\cite{bMCRB} (see Lemma 1 and equations (23) and (24) in~\cite{bMCRB}). For a misspecified-unbiased (MS-unbiased) estimator ${\boldsymbol{\hat \eta }}\left( {{{\bf{y}}_s}} \right)$, its mean equals ${{\boldsymbol{\eta }}_0}$.

\subsubsection{MCRB Derivation}
\label{MCRBsensing}
MCRB is defined as the lower bound on the error covariance matrix of any MS-unbiased estimator~\cite{bMCRB}, where the error refers to the difference between the estimator ${\boldsymbol{\hat \eta }}\left( {{{\bf{y}}_s}} \right)$ and the pseudo-true parameters ${{\boldsymbol{\eta }}_0}$, namely
\begin{equation}\label{E_MCRB}
\mathbb{E}\left\{ {\left( {{\boldsymbol{\hat \eta }}\left( {{{\bf{y}}_s}} \right) - {{\boldsymbol{\eta }}_0}} \right){{\left( {{\boldsymbol{\hat \eta }}\left( {{{\bf{y}}_s}} \right) - {{\boldsymbol{\eta }}_0}} \right)}^{\rm{T}}}} \right\} \succeq \rm{MCRB}\left( {{{\boldsymbol{\eta }}_0}} \right).
\end{equation}
The MCRB is further computed as
\begin{equation}\label{MCRB}
\rm{MCRB} \buildrel \Delta \over = {\bf{A}}_{{{\boldsymbol{\eta }}_0}}^{ - 1}{{\bf{B}}_{{{\boldsymbol{\eta }}_0}}}{\bf{A}}_{{{\boldsymbol{\eta }}_0}}^{ - 1}.
\end{equation}
The $\left( {i,j} \right)$-th element of ${{\bf{A}}_{{{\boldsymbol{\eta }}_0}}}\in {\mathbb{R}^{4\times 4}}$ and ${{\bf{B}}_{{{\boldsymbol{\eta }}_0}}}\in {\mathbb{R}^{4\times 4}}$ are given in (\ref{MCRBA}) and (\ref{MCRBB}), respectively, where ${\boldsymbol{\varepsilon }}\left( {\boldsymbol{\eta }} \right) = {\boldsymbol{\bar \mu }} - {\boldsymbol{\tilde \mu }}\left( {\boldsymbol{\eta }} \right)$.
\begin{figure*}[!t]
\begin{equation}\label{MCRBA}
{\left[ {{{\bf{A}}_{{{\boldsymbol{\eta }}_0}}}} \right]_{ij}} = \mathbb{E} \left\{ {{{\left. {\frac{{{\partial ^2}}}{{\partial {\eta _i}\partial {\eta _j}}}\log \tilde p\left( {{{\bf{y}}_s}\left| {\boldsymbol{\eta }} \right.} \right)} \right|}_{{\boldsymbol{\eta }} = {{\boldsymbol{\eta }}_0}}}} \right\} = {\left. {\frac{2}{{\sigma _s^2}}{\rm{Re}}\left[ {{\boldsymbol{\varepsilon }}{{\left( {\boldsymbol{\eta }} \right)}^{\rm{H}}}\frac{{{\partial ^2}{\boldsymbol{\tilde \mu }}\left( {\boldsymbol{\eta }} \right)}}{{\partial {\eta _i}\partial {\eta _j}}} - {{\left( {\frac{{\partial {\boldsymbol{\tilde \mu }}\left( {\boldsymbol{\eta }} \right)}}{{\partial {\eta _i}}}} \right)}^{\rm{H}}}\frac{{\partial {\boldsymbol{\tilde \mu }}\left( {\boldsymbol{\eta }} \right)}}{{\partial {\eta _j}}}} \right]} \right|_{{\boldsymbol{\eta }} = {{\boldsymbol{\eta }}_0}}}.
\end{equation}
\vspace{-0.7cm}
\end{figure*}
\begin{figure*}[!t]
\begin{equation}\label{MCRBB}
{\left[ {{{\bf{B}}_{{{\boldsymbol{\eta }}_0}}}} \right]_{ij}} = {\left. {\left[ \mathbb{E}{\left\{ {\frac{{\partial \log \tilde p\left( {{{\bf{y}}_s}\left| {\boldsymbol{\eta }} \right.} \right)}}{{\partial {\eta _i}}}\frac{{\partial \log \tilde p\left( {{{\bf{y}}_s}\left| {\boldsymbol{\eta }} \right.} \right)}}{{\partial {\eta _j}}}} \right\} - \frac{{\partial {D_{KL}}}}{{\partial {\eta _i}}}\frac{{\partial {D_{KL}}}}{{\partial {\eta _j}}}} \right]} \right|_{{\boldsymbol{\eta }} = {{\boldsymbol{\eta }}_0}}} = \frac{2}{{\sigma _s^2}}{\left. {{\rm{Re}}\left\{ {{{\left( {\frac{{\partial {\boldsymbol{\tilde \mu }}\left( {\boldsymbol{\eta }} \right)}}{{\partial {\eta _i}}}} \right)}^{\rm{H}}}\frac{{\partial {\boldsymbol{\tilde \mu }}\left( {\boldsymbol{\eta }} \right)}}{{\partial {\eta _j}}}} \right\}} \right|_{{\boldsymbol{\eta }} = {{\boldsymbol{\eta }}_0}}}.
\end{equation}
\vspace{-0.6cm}
\end{figure*}
This work focuses on AoD estimation ${\boldsymbol{\phi}} = {\left[ {{\phi _e},{\phi _a}} \right]^{\rm{T}}}$. The corresponding ${\rm{MCRB}}_{\boldsymbol{\phi}}$ is obtained from ${{\bf{A}}_{{{\boldsymbol{\eta }}_0}}}$ and ${{\bf{B}}_{{{\boldsymbol{\eta }}_0}}}$, which consist of block matrices w.r.t the parameters $\boldsymbol{\phi}$ and $\alpha$, namely ${{\bf{A}}_{\boldsymbol{\phi} \boldsymbol{\phi} }},{{\bf{A}}_{\alpha \boldsymbol{\phi} }},{{\bf{A}}_{\boldsymbol{\phi} \alpha }},{\bf{A}}_{\alpha \alpha },{{\bf{B}}_{\boldsymbol{\phi} \boldsymbol{\phi} }},{{\bf{B}}_{\alpha \boldsymbol{\phi} }},{{\bf{B}}_{\boldsymbol{\phi} \alpha }},$ ${\bf{B}}_{\alpha \alpha }$, whose derivations are omitted due to limited space\footnote{The derivations of ${{\bf{A}}_{\boldsymbol{\phi} \boldsymbol{\phi} }},{{\bf{A}}_{\alpha \boldsymbol{\phi} }},{{\bf{A}}_{\boldsymbol{\phi} \alpha }},{\bf{A}}_{\alpha \alpha },{{\bf{B}}_{\boldsymbol{\phi} \boldsymbol{\phi} }},{{\bf{B}}_{\alpha \boldsymbol{\phi} }},{{\bf{B}}_{\boldsymbol{\phi} \alpha }},$ ${\bf{B}}_{\alpha \alpha }$ follow steps similar to those in Appendix A of~\cite{bLW}, but are adapted to holographic beamforming in this paper. In particular, \cite{bLW} focuses on complex-valued RIS phase shifts, whereas we incorporate the practical constraint on RHS elements for holographic beamforming, where only their real-valued amplitudes can be controlled.}.
%are given in Appendix~\ref{AppCalcuAB}.
The inverse of the block matrix ${{\bf{A}}_{{{\boldsymbol{\eta }}_0}}}$ is
\begin{equation}\label{Ainv}
{\bf{A}}_{{{\boldsymbol{\eta }}_0}}^{ - 1} \!=\! \left[ \! {\begin{array}{*{20}{c}}
{{{\bf{Z}}^{ - 1}}}&{ - {{\bf{Z}}^{ - 1}}{{\bf{A}}_{\alpha \phi }}{\bf{A}}_{\alpha \alpha }^{ - 1}}\\
{ - {\bf{A}}_{\alpha \alpha }^{ - 1}{{\bf{A}}_{\phi \alpha }}{{\bf{Z}}^{ - 1}}}&{{\bf{A}}_{\alpha \alpha }^{ - 1} + {\bf{A}}_{\alpha \alpha }^{ - 1}{{\bf{A}}_{\phi \alpha }}{{\bf{Z}}^{ - 1}}{{\bf{A}}_{\alpha \phi }}{\bf{A}}_{\alpha \alpha }^{ - 1}}
\end{array}} \! \right],
\end{equation}
where ${\bf{Z}} = {{\bf{A}}_{\boldsymbol{\phi} \boldsymbol{\phi} }} - {{\bf{A}}_{\alpha \boldsymbol{\phi} }}{\bf{A}}_{\alpha \alpha }^{ - 1}{{\bf{A}}_{\boldsymbol{\phi} \alpha }}\in \mathbb{R}^{2 \times 2}$. Substituting ${\bf{A}}_{{{\boldsymbol{\eta }}_0}}^{ - 1}$ and ${\bf{B}}_{{{\boldsymbol{\eta }}_0}}$ into (\ref{MCRB}) yields ${\rm{MCRB}}$, whose first diagonal block matrix is namely ${{\rm{MCRB}}_{\boldsymbol{\phi}} }$, obtained as \eqref{MCRB_AoD}. We leverage ${\rm{MCRB}}_{\boldsymbol{\phi}}$ to characterize the lower bound on the mean square error of AoD estimation under faulty RHS elements.

\begin{figure*}[!t]
\begin{equation}\label{MCRB_AoD}
{{\rm{MCRB}}_\phi } = {{\bf{Z}}^{ - 1}}\left( {{{\bf{B}}_{\phi \phi }} - {{\bf{B}}_{\alpha \phi }}{\bf{B}}_{\alpha \alpha }^{ - 1}{{\bf{B}}_{\phi \alpha }} + \left( {{{\bf{A}}_{\alpha \phi }} + {{\bf{B}}_{\alpha \phi }}} \right){\bf{B}}_{\alpha \alpha }^{ - 1}\left( {{{\bf{A}}_{\phi \alpha }} + {{\bf{B}}_{\phi \alpha }}} \right)} \right){{\bf{Z}}^{ - 1}} = {{\bf{Z}}^{ - 1}}{\bf{U}}{{\bf{Z}}^{ - 1}}.
\end{equation}
\vspace{-0.7cm}
\end{figure*}

\begin{figure*}[t]
\noindent\rule{\linewidth}{0.5pt}
\vspace{-0.8cm}
\end{figure*}
%%%%%%%%%%%%%%%%%%%%%%%%

%their amplitudes are unknown. We assume that the locations of faulty elements are known for the subsequent optimization\footnote{\color{blue}The locations of faulty elements can be obtained using the method proposed in \cite{bfaultyDetect1,bfaultyDetect2}. The case where the locations of faulty elements are unknown is left for future work.}
\section{Problem Formulation and Solution}
In Sections~\ref{COMSINR} and~\ref{MCRBsensing}, we derived the metrics SINR and MCRB to quantify the impact of faulty RHS elements. In this section, we focus on mitigating the performance degradation caused by such faulty elements in both sensing and communication. To this end, we jointly design digital and holographic beamforming. Specifically, we formulate an optimization problem that minimizes the MCRBs of ${\phi _e}$ and ${\phi _a}$, subject to constraints on the communication SINR, the transmit power, and the functioning RHS amplitudes, namely
\begin{subequations}\label{OptP1}
\begin{align}
& ({\rm P}1) \mathop {\min }\limits_{\{ {\bf{W}},{{\bf{R}}_s},{{\bar \psi }_n}\} } {\rm{tr}}\left( {{\rm{MCRB}}_\phi } \right) = {\rm{tr}}\left( {{{\bf{Z}}^{ - 1}}{\bf{U}}{{\bf{Z}}^{ - 1}}} \right) \label{P1Obj}\\
\mbox{s.t.}&
\begin{array}{*{20}{c}}
{\frac{{{{\left| {{\bf{h}}_k^{\rm{H}}{\bf{\bar \Psi Q}}{{\bf{w}}_k}} \right|}^2}}}{{\sum\limits_{i \ne k}^K {{{\left| {{\bf{h}}_k^{\rm{H}}{\bf{\bar \Psi Q}}{{\bf{w}}_i}} \right|}^2}}  + {\bf{h}}_k^{\rm{H}}{\bf{\bar \Psi Q}}{{\bf{R}}_s}{{\bf{Q}}^{\rm{H}}}{{{\bf{\bar \Psi }}}^{\rm{H}}}{{\bf{h}}_k} + \sigma _k^2}}}\\
{ \ge {\gamma _{th}},\forall k \in {\mathcal K}}
\end{array} \label{const_COM} \\
&{\rm{tr}}\left( {{\bf{\bar \Psi Q}}\left( {{\bf{W}}{{\bf{W}}^{\rm{H}}} + {{\bf{R}}_s}} \right){{\bf{Q}}^{\rm{H}}}}{{\bf{\bar \Psi}} ^{\rm{H}}} \right) \le {P_{\max }},\label{const_Pwr} \\
& {{\bf{R}}_s} \succeq 0, \label{const_Rs} \\
& {{\bar \psi } _n} \in [0,1], \forall n \in {\mathcal S_W}. \label{const_RHS}
\end{align}
\end{subequations}
Among all constraints, \eqref{const_COM} guarantees that the SINR of each UE is no less than the predefined threshold ${\gamma _{th}}$. \eqref{const_Pwr} and \eqref{const_Rs} enforce that the total transmit power for hybrid beamforming allocated to sensing and communication does not exceed the maximum power $P_{\max }$ at the BS. \eqref{const_RHS} represents the amplitude constraint imposed on the set of functional RHS elements. Problem $\rm{P}$1 is highly non-convex due to the intricate structure of the objective function, strong coupling among variables, and nonlinear constraints.

Next, we propose a solution to address problem $\rm{P}$1. Due to the inverse operation and the multiplicative relationships among the terms in~(\ref{MCRB_AoD}), the variables are strongly coupled, making problem $\rm{P}$1 intractable to solve. To overcome this difficulty, we introduce two auxiliary matrices ${\bf{\tilde C}}\in \mathbb{R}^{2 \times 2} \succeq 0$ and ${\bf{\tilde D}}\in \mathbb{R}^{2 \times 2} \prec 0$, which facilitate problem reformulation and enable a more tractable solution. Specifically, let $( {{{\bf{B}}_{{\boldsymbol{\phi}} {\boldsymbol{\phi}} }} + \left( {{{\bf{A}}_{\alpha {\boldsymbol{\phi}} }} + {{\bf{B}}_{\alpha {\boldsymbol{\phi}} }}} \right){\bf{B}}_{\alpha \alpha }^{ - 1}\left( {{{\bf{A}}_{{\boldsymbol{\phi}} \alpha }} + {{\bf{B}}_{{\boldsymbol{\phi}} \alpha }}} \right)} ) \le {\bf{\tilde C}}$ and ${{\bf{Z}}^{ - 1}} = {( {{{\bf{A}}_{{\boldsymbol{\phi}} {\boldsymbol{\phi}} }} - {{\bf{A}}_{\alpha {\boldsymbol{\phi}} }}{\bf{A}}_{\alpha \alpha }^{ - 1}{{\bf{A}}_{{\boldsymbol{\phi}} \alpha }}})^{ - 1}} \ge {\bf{\tilde D}}$ in postive semidefinite (PSD) space, which are equivalent to \eqref{const_AuxilC} and \eqref{const_AuxilD}, respectively, based on Schur complement.
\begin{equation}\label{const_AuxilC}
\left[ {\begin{array}{*{20}{c}}
{{\bf{\tilde C}} -  {{\bf{B}}_{{\boldsymbol{\phi}} {\boldsymbol{\phi}}}}}&{{{\bf{A}}_{\alpha {\boldsymbol{\phi}} }} + {{\bf{B}}_{\alpha {\boldsymbol{\phi}} }}}\\
{{{\bf{A}}_{{\boldsymbol{\phi}} \alpha }} + {{\bf{B}}_{{\boldsymbol{\phi}} \alpha }}}&{{{\bf{B}}_{\alpha \alpha }}}
\end{array}} \right] \succeq 0.
\end{equation}
\begin{equation}\label{const_AuxilD}
- \left[ {\begin{array}{*{20}{c}}
{{\bf{\tilde D}}}&{{{\bf{I}}_2}}&{\bf{0}}\\
{{{\bf{I}}_2}}&{{{\bf{A}}_{{\boldsymbol{\phi}} {\boldsymbol{\phi}} }}}&{{{\bf{A}}_{\alpha {\boldsymbol{\phi}} }}}\\
{\bf{0}}&{{{\bf{A}}_{{\boldsymbol{\phi}} \alpha }}}&{{{\bf{A}}_{\alpha \alpha }}}
\end{array}} \right] \succeq 0.
\end{equation}
On the condition of \eqref{const_AuxilC} and \eqref{const_AuxilD}, we obtain the relation ${\rm{tr}}( {{{\bf{Z}}^{ - 1}}{\bf{U}}{{\bf{Z}}^{ - 1}}})$ $\le$ ${\rm{tr}}( {{{\bf{Z}}^{ - 1}}( {{\bf{\tilde C}} - {{\bf{B}}_{\alpha {\boldsymbol{\phi}}}}{\bf{B}}_{\alpha \alpha }^{ - 1}{{\bf{B}}_{{\boldsymbol{\phi}} \alpha }}}){{\bf{Z}}^{ - 1}}} )\le$ ${\rm{tr}}( {{{\bf{\tilde D}}}( {{\bf{\tilde C}} - {{\bf{B}}_{\alpha {\boldsymbol{\phi}}}}{\bf{B}}_{\alpha \alpha }^{ - 1}{{\bf{B}}_{{\boldsymbol{\phi}} \alpha }}}){{\bf{\tilde D}}}} )$ $\le$ $ {\rm{tr}}({{\bf{\tilde C}} - {{\bf{B}}_{\alpha {\boldsymbol{\phi}}}}{\bf{B}}_{\alpha \alpha }^{ - 1}{{\bf{B}}_{{\boldsymbol{\phi}} \alpha }}}){\rm{tr}}({{\bf{\tilde D\tilde D}}})$ since ${\bf{\tilde C}} - {{\bf{B}}_{\alpha {\boldsymbol{\phi}}}}{\bf{B}}_{\alpha \alpha }^{ - 1}{{\bf{B}}_{{\boldsymbol{\phi}} \alpha }}$ is PSD and ${\bf{\tilde D\tilde D}}$ is positive definite. The proof of the aforementioned transformation is omitted due to limited space.
Thus, subject to \eqref{const_AuxilC} and \eqref{const_AuxilD}, ${\rm{tr}}( {{\bf{\tilde C}} - {{\bf{B}}_{\alpha {\boldsymbol{\phi}}}}{\bf{B}}_{\alpha \alpha }^{ - 1}{{\bf{B}}_{{\boldsymbol{\phi}}\alpha }}}){\rm{tr}}( {{\bf{\tilde D\tilde D}}})$ is an upper bound of the original objective ${\rm{tr}}({{{\bf{Z}}^{ - 1}}{\bf{U}}{{\bf{Z}}^{ - 1}}})$, and problem $\rm{P}$1 is relaxed into problem $\rm{P}$1.1, given by
\begin{subequations}\label{OptP11}
\begin{align}
({\rm P}1.1) & \mathop {\min }\limits_{\{ {\bf{W}},{{\bf{R}}_s},{{\bar \psi }_n},{\bf{\tilde C}},{\bf{\tilde D}}\} } {\rm{tr}}\left( {{\bf{\tilde C}} - {{\bf{B}}_{\alpha {\boldsymbol{\phi}} }}{\bf{B}}_{\alpha \alpha }^{ - 1}{{\bf{B}}_{{\boldsymbol{\phi}} \alpha }}} \right){\rm{tr}}\left( {{{\bf{\tilde D}}}{{\bf{\tilde D}}}} \right)\label{P11Obj}\\
\mbox{s.t.}&
\eqref{const_COM},\eqref{const_Pwr},\eqref{const_Rs},\eqref{const_RHS},\eqref{const_AuxilC},\eqref{const_AuxilD}.\\
& {{\bf{\tilde C}}} \succeq 0, {{\bf{\tilde D}}} \prec 0.\label{const_CDPSD}
\end{align}
\end{subequations}
In problem $\rm{P}$1.1, the degree of coupling in the objective function is significantly reduced compared to the original objective ${{\rm{MCRB}}_{\boldsymbol{\phi}}}$ (\ref{MCRB_AoD}). However, $\{ {\bf{W}},{{\bf{R}}_s},{{\bar \psi }_n}\}$ are still coupled in the objective function (\ref{P11Obj}) and constraints (\ref{const_COM}), (\ref{const_AuxilC}), (\ref{const_AuxilD}). $\{{\bf{\tilde C}},{\bf{\tilde D}}\}$ are coupled in the objective (\ref{P11Obj}). To address this issue, we partition these variables into two blocks, namely $\{ {\bf{W}},{{\bf{R}}_s},{\bf{\tilde C}}\}$ and $\{{{\bar \psi }_n},{\bf{\tilde D}}\}$, and then apply the BCD framework to iteratively update each block while keeping the variables in the other block fixed until convergence. The detailed solutions for each block are presented in the following.

\subsection{Sub-problem with Respect to $\{ {\bf{W}},{{\bf{R}}_s},{\bf{\tilde C}}\}$}
The sub-problem w.r.t. $\{ {\bf{W}},{{\bf{R}}_s},{\bf{\tilde C}}\}$ is written as follows:
\begin{subequations}\label{OptP12}
\begin{align}
({\rm P}1.2) & \mathop {\min }\limits_{\{ {\bf{W}},{{\bf{R}}_s},{\bf{\tilde C}}\} } {\rm{tr}}\left( {{\bf{\tilde C}} - {{\bf{B}}_{\alpha {\boldsymbol{\phi}} }}{\bf{B}}_{\alpha \alpha }^{ - 1}{{\bf{B}}_{{\boldsymbol{\phi}} \alpha }}} \right)\label{P12Obj}\\
\mbox{s.t.}&
\eqref{const_AuxilC},\eqref{const_AuxilD},\eqref{const_CDPSD}, \eqref{const_COM},\eqref{const_Pwr},\eqref{const_Rs}.
\end{align}
\end{subequations}
For UE $k\in {\mathcal K}$, we define ${{\bf{H}}_k} = {( {{\bf{h}}_k^{\rm{H}}{\bf{\bar \Psi Q}}} )^{\rm{H}}}( {{\bf{h}}_k^{\rm{H}}{\bf{\bar \Psi Q}}} ) \in {\mathbb{C}^{N_{RF} \times N_{RF}}}$ and ${{\bf{W}}_k} = {{\bf{w}}_k}{\bf{w}}_k^{\rm{H}} \in {\mathbb{C}^{N_{RF} \times N_{RF}}}$, satisfying ${{\bf{W}}_k} \succeq 0$ and ${\rm{rank}}\left( {\bf{W}}_k \right) = 1$. The SINR constraint (\ref{const_COM}) for UE $k$ can be transformed into the following form.
\begin{equation}\label{SINRtransW}
\left( {\frac{1}{{{\gamma _{th}}}} \!+\! 1} \right){\rm{tr}}\left( {{{{\bf{H}}}_k}{{\bf{W}}_k}} \right) \!-\! {\rm{tr}}\left( {{{{\bf{H}}}_k}\left( {\sum\limits_{k = 1}^K {{{\bf{W}}_k} \!+\! {{\bf{w}}_s}{\bf{w}}_s^{\rm{H}}} } \right)} \right) \!\ge\! \sigma _k^2.
\end{equation}
To address the variables coupling of ${{{\bf{B}}_{\alpha {\boldsymbol{\phi}} }}{\bf{B}}_{\alpha \alpha }^{ - 1}{{\bf{B}}_{{\boldsymbol{\phi}} \alpha }}}$ in \eqref{P12Obj}, we apply the MM technique, where ${\rm{tr}}\left( {{{\bf{B}}_{\alpha {\boldsymbol{\phi}} }}{\bf{B}}_{\alpha \alpha }^{ - 1}{{\bf{B}}_{{\boldsymbol{\phi}} \alpha }}} \right)$ is linearized by its first-order Taylor approximation, resulting in a tractable lower bound ${\rm{tr}}({{\bf{B'}}})= 2{\rm{tr}}( {\mathring{{\bf{B}}_{\alpha {\boldsymbol{\phi}} }} {{\mathring{( {{\bf{B}}_{\alpha \alpha }^{ - 1}})}} }{{\bf{B}}_{{\boldsymbol{\phi}} \alpha }}}) \!-\! {\rm{tr}}( {{{\mathring{({{\bf{B}}_{\alpha \alpha }^{ - 1}})}}}{\mathring{\bf{B}}_{{\boldsymbol{\phi}} \alpha }} {\mathring{\bf{B}}_{\alpha {\boldsymbol{\phi}} }} {\mathring{{( {{\bf{B}}_{\alpha \alpha }^{ - 1}} )}}}})$. %~\cite{MM}.
The values with $\mathring {}$ denote the updated values from the previous iteration. Thus, the upper bound of \eqref{P12Obj} is obtained as ${\rm{tr}}({{\bf{\tilde C}}}) - {\rm{tr}}({{\bf{B'}}})$. The problem to be solved at each MM iteration is given as follows.
\begin{subequations}\label{OptP13}
\begin{align}
&({\rm P}1.3) \mathop {\min }\limits_{\{ {\bf{W}}_k,{{\bf{R}}_s},{\bf{\tilde C}}\} } \left( {{\rm{tr}}\left( {{\bf{\tilde C}}} \right) - {\rm{tr}}\left( {{\bf{B'}}({{\bf{R}}_x})} \right)} \right) \label{P13Obj}\\
\mbox{s.t.}&
{\rm{tr}}\left( {{\bf{\bar \Psi Q}}\left( {\sum\nolimits_{k = 1}^K {{{\bf{W}}_k} + {{\bf{R}}_s}} } \right){{\bf{Q}}^{\rm{H}}}{{{\bf{\bar \Psi }}}^{\rm{H}}}} \right) \le {P_{\max}}, \label{const_Pwr1}\\
& {{\bf{W}}_k} \succeq 0, {\rm{rank}}\left( {\bf{W}}_k \right) = 1, \forall k \in {\mathcal K}, \label{const_WkPSD} \\
&\eqref{const_Rs},\eqref{const_AuxilC},\eqref{const_AuxilD},\eqref{const_CDPSD},\eqref{SINRtransW}.
\end{align}
\end{subequations}
Problem $\rm{P}$1.3 is non-convex due to the rank-one constraint, which can be dropped and result in problem $\rm{P}$1.3 a semidefinite programming (SDP) problem, denoted as problem ($\rm{P}$1.3 SDP). Note that the optimal ${\bf{W}}_k$ obtained from problem ($\rm{P}$1.3 SDP) may have a higher rank. The optimal solution to problem $\rm{P}$1.3 satisfying the rank-one constraint (\ref{const_WkPSD}) can be recovered by applying (26) and (27) in~\cite{bXXS}.

%%%%%%%%%%%%%%%%%%%%%%%%% Sub-prob2 %%%%%%%%%%%%%%%%%%%%%%%%%%
\subsection{Sub-problem with Respect to $\{{\boldsymbol{\bar \psi}},{\bf{\tilde D}}\}$}
The sub-problem w.r.t. $\{{\boldsymbol{\bar \psi}},{\bf{\tilde D}}\}$ is shown as follows:
\begin{subequations}\label{OptP14}
\begin{align}
({\rm P}1.4)\; & \mathop {\min }\limits_{\{{\boldsymbol{\bar \psi}},{\bf{\tilde D}}\} }{\rm{tr}}\left( {{\bf{\tilde D\tilde D}}} \right) \label{P14Obj}\\
\mbox{s.t.}& \;
\eqref{const_COM}, \eqref{const_RHS},\eqref{const_AuxilC}, \eqref{const_AuxilD}, \eqref{const_CDPSD}.
\end{align}
\end{subequations}
Since $\bf{\tilde D}$ is symmetric and real, ${\rm{tr}}( {{\bf{\tilde D\tilde D}}}) = {\rm{tr}}( {{{{\bf{\tilde D}}}^{\rm{T}}}{\bf{\tilde D}}}){\rm{ = }}\| {{\bf{\tilde D}}}\|_F^2$ is obtained. The squared Frobenius norm is a sum of the squares of all entries in matrix ${\bf{\tilde D}}$. Thus, \eqref{P14Obj} is convex w.r.t. ${\bf{\tilde D}}$. Additionally, since ${\boldsymbol {\bar \Psi}}  = {\rm{diag}}({\boldsymbol{\bar \psi}}^{\rm{H}})$, we transform all the involved constraints w.r.t. ${\boldsymbol {\bar \Psi}}$ into an equivalent form w.r.t. ${\boldsymbol{\bar \psi}}$ to facilitate solving RHS amplitude variables, where $W$ functional elements from ${\mathcal S_W}$ out of $N_t$ elements are design variables. To be specific, for \eqref{const_Pwr}, we operate the transformation ${\rm{tr}}( {{\rm{diag}}({{\boldsymbol{\bar \psi }}^{\rm{H}}}){\bf{Q}}( {{\bf{W}}{{\bf{W}}^{\rm{H}}} + {{\bf{R}}_s}} ){{\bf{Q}}^{\rm{H}}}{\rm{diag}}({\boldsymbol{\bar \psi }})} ) = {{\boldsymbol{\bar \psi}}^{\rm{H}}}({{\bf{Q}}({{\bf{W}}{{\bf{W}}^{\rm{H}}} + {{\bf{R}}_s}}){{\bf{Q}}^{\rm{H}}} \odot {{\bf{I}}_{{N_t}}}}){\boldsymbol{\bar \psi}}$~\cite{bMatrix}, resulting in
\begin{equation}\label{const_Pwr2}
{{\boldsymbol{\bar \psi}}^{\rm{H}}}\left( {{\bf{Q}}\left( {{\bf{W}}{{\bf{W}}^{\rm{H}}} + {{\bf{R}}_s}} \right){{\bf{Q}}^{\rm{H}}} \odot {{\bf{I}}_{{N_t}}}} \right){\boldsymbol{\bar \psi}} \le {P_{\max }}.
\end{equation}
where ${{\bf{I}}_{{N_t}}}$ is the identity matrix. Meanwhile, for the sensing related constraints \eqref{const_AuxilC} and \eqref{const_AuxilD}, the variable ${\boldsymbol {\bar \Psi}}$ affects ${{\bf{A}}_{{\boldsymbol{\phi}} {\boldsymbol{\phi}} }}$, ${{\bf{A}}_{\alpha {\boldsymbol{\phi}} }}$, ${\bf{A}}_{{\boldsymbol{\phi}} \alpha }$ under the faulty RHS case.
%, as shown in \eqref{Fbarsub}, \eqref{Aphiphi}, \eqref{Aphialpha}, and \eqref{Aalphaphi}.
Likewise, we transform ${{\bf{A}}_{\boldsymbol{\phi} \boldsymbol{\phi} }},{{\bf{A}}_{\alpha \boldsymbol{\phi} }},{{\bf{A}}_{\boldsymbol{\phi} \alpha }}$ into the expressions w.r.t. ${\boldsymbol{\bar \psi}}$. Let us take the first element in ${{\bf{A}}_{\boldsymbol{\phi} \boldsymbol{\phi} }}$ as an example, which can be reconstructed as follows~\cite{bMatrix}.
\begin{equation}\label{FbarTrans}
\begin{array}{l}
{\rm{tr}}\left( {{{{\bf{\ddot \Omega }}}_{{\phi _e}{\phi _e}}} {\bf{\tilde \Psi Q}}\left( {{\bf{W}}{{\bf{W}}^{\rm{H}}} + {{\bf{R}}_s}} \right){{\bf{Q}}^{\rm{H}}}{{\bf{\bar \Psi }}^{\rm{H}}} {{\bf{\Omega }}^{\rm{H}}}} \right)\\
= {\rm{tr}}\left( {{{{\bf{\ddot \Omega }}}_{{\phi _e}{\phi _e}}}{\rm{diag}}({{{\boldsymbol{\tilde \psi }}}^{\rm{H}}}){\bf{Q}}\left( {{\bf{W}}{{\bf{W}}^{\rm{H}}} + {{\bf{R}}_s}} \right){{\bf{Q}}^{\rm{H}}}{\rm{diag}}({\boldsymbol{\bar \psi }}){{\bf{\Omega }}^{\rm{H}}}} \right)\\
= {{{\boldsymbol{\tilde \psi }}}^{\rm{H}}}\left( {{\bf{Q}}\left( {{\bf{W}}{{\bf{W}}^{\rm{H}}} + {{\bf{R}}_s}} \right){{\bf{Q}}^{\rm{H}}} \odot \left( {{{\bf{\Omega }}^{\rm{H}}}{{{\bf{\ddot \Omega }}}_{{\phi _e}{\phi _e}}}} \right)} \right){\boldsymbol{\bar \psi}}.
\end{array}
\end{equation}
Other elements in ${{\bf{A}}_{\boldsymbol{\phi} \boldsymbol{\phi} }},{{\bf{A}}_{\alpha \boldsymbol{\phi} }},{{\bf{A}}_{\boldsymbol{\phi} \alpha }}$ are transformed similarly, with details omitted due to limited space. The SINR constraint \eqref{const_COM} can also be transformed as ${{\boldsymbol{\bar \psi }}}^{\rm{T}}{{\bf{V}}_k}{\boldsymbol{\bar \psi }} \ge {\gamma _{th}}\sigma _k^2, \forall k \in {\mathcal K}$, where ${{\bf{V}}_k}$ is given by
{\setlength{\abovedisplayskip}{2.5pt}
\setlength{\belowdisplayskip}{2.5pt}
\begin{equation}\label{Vk}
{{{\bf{V}}_k} \!=\! {\bf{Q}}\left( {{{\bf{W}}_k} \!-\! {\gamma _{th}}\left( {\sum\limits_{i \ne k}^K {{{\bf{W}}_i}} \!+\! {{\bf{R}}_s}} \right)} \right){{\bf{Q}}^{\rm{H}}} \odot \left( {{{\bf{h}}_k}{\bf{h}}_k^{\rm{H}}} \right)^{\rm{T}}}.
\end{equation}}
Next, we define an auxiliary variable ${\bf{\Phi }} = {\boldsymbol{\bar \psi}}{{\boldsymbol{\bar \psi}}^{\rm{T}}}$, satisfying ${\bf{\Phi}} \succeq 0$ and ${\rm{rank}}( {\bf{\Phi}} ) = 1$, which is equivalent to the following constraint using Schur complement, on the condition that ${\rm{rank}}( {\bf{\Phi}} ) = 1$ is satisfied.
\begin{equation}\label{PhiWPhiWschur}
\left[ {\begin{array}{*{20}{c}}
{\bf{\Phi }}&{\boldsymbol{\bar \psi}}\\
{{\boldsymbol{\bar \psi}}^{\rm{T}}}&1
\end{array}} \right] \succeq 0.
\end{equation}
The SINR constraint is further converted into
\begin{equation}\label{SINRtransV}
{\rm{tr}}\left( {{{\bf{V}}_k}{\bf{\Phi}}} \right) \ge {\gamma _{th}}\sigma _k^2,\forall k \in {\cal K}.
\end{equation}
Thus, the problem to be solved is rewritten as problem $\rm{P}$1.5.
{\setlength{\abovedisplayskip}{2.5pt}
\setlength{\belowdisplayskip}{2.5pt}
\begin{subequations}\label{OptP15}
\begin{align}
&({\rm P}1.5) \mathop {\min }\limits_{\{ {\boldsymbol{\bar \psi}},{\bf{\Phi}}, {\bf{\tilde D}}\} } {\rm{tr}}\left( {{\bf{\tilde D\tilde D}}} \right) \label{P15Obj}\\
\mbox{s.t.}&
\left[ {\begin{array}{*{20}{c}}
{{\bf{\tilde C}} - {{\bf{B}}_{{\boldsymbol{\phi}} {\boldsymbol{\phi}} }}}&{{{\bf{A}}_{\alpha {\boldsymbol{\phi}} }}({\boldsymbol{\bar \psi}}) + {{\bf{B}}_{\alpha {\boldsymbol{\phi}} }}}\\
{{{\bf{A}}_{{\boldsymbol{\phi}} \alpha }}({\boldsymbol{\bar \psi}}) + {{\bf{B}}_{{\boldsymbol{\phi}} \alpha }}}&{{{\bf{B}}_{\alpha \alpha }}}
\end{array}} \right] \succeq 0, \label{const_AuxilC4} \\
& - \left[ {\begin{array}{*{20}{c}}
{{\bf{\tilde D}}}&{{{\bf{I}}_2}}&{\bf{0}}\\
{{{\bf{I}}_2}}&{{{\bf{A}}_{{\boldsymbol{\phi}} {\boldsymbol{\phi}} }}({\boldsymbol{\bar \psi}})}&{{{\bf{A}}_{\alpha {\boldsymbol{\phi}} }}({\boldsymbol{\bar \psi}})}\\
{\bf{0}}&{{{\bf{A}}_{{\boldsymbol{\phi}} \alpha }}({\boldsymbol{\bar \psi}})}&{{\bf{A}}_{\alpha \alpha }}
\end{array}} \right] \succeq 0 , \label{const_AuxilD4} \\
&{\rm{rank}}( {\bf{\Phi}} ) = 1,{\bf{\Phi}} \succeq 0, \label{const_rank1} \\
&\eqref{const_RHS},\eqref{const_Pwr2},\eqref{PhiWPhiWschur},\eqref{SINRtransV}.
\end{align}
\end{subequations}}
In problem $\rm{P}$1.5, only the rank-one constraint \eqref{const_rank1} is not convex. To address this issue, we apply the SCA technique and the penalty-based method to ensure the rank-one constraint \eqref{const_rank1}, with similar operations found in~\cite{bLW}.

%%%%%%%%%%%%%%%%%%% Summarization %%%%%%%%%%%%%%%%%%%%%%%
Algorithm~\ref{AlgoP1} summarizes the solution to $\rm{P}$1, where the problem is decomposed into two sub-problems and solved alternately using the BCD framework until convergence.

%%%%%%%%%%%%%%%%%%% Convergence Analysis %%%%%%%%%%%%%%%%%%%%%%%
\textbf{Convergence Analysis}:
Both the MM-based and SCA-based sub-problems construct convex surrogates that yield monotonically tightening upper bounds, ensuring convergence to a stationary point. Under alternating optimization, the objective value forms a monotonically non-increasing sequence, thereby guaranteeing convergence to a stationary point.

%%%%%%%%%%%%%%%%%%% Complexity Analysis %%%%%%%%%%%%%%%%%%%%%%%
\textbf{Complexity Analysis}:
The computational cost arises from updating $\{ {\bf{W}},{{\bf{R}}_s},{\bf{\tilde C}}\}$ and $\{{\boldsymbol{\bar \psi}},{\bf{\tilde D}}\}$. Assuming the interior-point method is applied for solving SDP problems in CVX, their complexities are ${\mathcal{O}\left( {{K^3}N_{RF}^6{I_{{\rm{MM}}}}} \right)}$ and ${\mathcal{O}\left( {{W^{6}}{I_{{\rm{SCA}}}}} \right)}$. Thus, the overall complexity of solving problem ${\rm P}1$ is $\mathcal{O}( {\left( {{K^3}N_{RF}^6{I_{{\rm{MM}}}} + {W^{6}}{I_{{\rm{SCA}}}}} \right){I_{{\rm{BCD}}}}} )$, where ${I_{{\rm{MM}}}}$, ${I_{{\rm{SCA}}}}$ and ${I_{{\rm{BCD}}}}$ denote the iteration numbers of MM, SCA, and BCD methods, respectively.

\begin{algorithm}[htbp]
\caption{BCD-based algorithm for solving problem ${\rm P}1$}
\label{AlgoP1}
\begin{algorithmic}[1]
\STATE \textbf{Initialization}: Initialize ${\boldsymbol{\bar \psi}}$, ${\bf{\tilde D}}$, ${\mathring{\bf{B}}_{\alpha {\boldsymbol{\phi}} }}$, ${\mathring{\bf{B}}_{{\boldsymbol{\phi}} \alpha }}$, and ${\mathring{ {{\bf{B}}_{\alpha \alpha }^{ - 1}}}}$ with feasible values.
\REPEAT 
\STATE Solve problem ${\rm P}1.2$ to obtain $\{ {\bf{W}},{{\bf{R}}_s},{\bf{\tilde C}}\}$.
\STATE Solve problem ${\rm P}1.4$ to obtain $\{{\boldsymbol{\bar \psi}},{\bf{\tilde D}}\}$.
\UNTIL the reduction of the objective function \eqref{P11Obj} is smaller than the threshold $\varepsilon $. \\
\STATE \textbf{Output}: $\{ {\bf{W}},{{\bf{R}}_s},{\boldsymbol{\bar \psi}},{\bf{\tilde C}},{\bf{\tilde D}}\}$. \\
\end{algorithmic}
\end{algorithm}

\addtolength{\topmargin}{0.1in}
\vspace{-0.1in}
\section{Performance Evaluation}
This section evaluates the performance of the proposed solution compared against three benchmarks: \textbf{i) Perfect RHS-lower bound (LB)}: All RHS elements operate without failures. The resulting performance is measured by the CRB, serving as a lower bound for practical faulty scenarios. \textbf{ii) Faulty RHS-naive}: Faulty elements are simply neglected. The amplitudes of functional elements are designed under the incorrect assumption of fully functional elements. \textbf{iii) Faulty RHS-random}: The functional elements are configured with random amplitudes to evaluate the performance gain achieved by optimizing the functional elements in our approach.

Simulations are performed in a three-dimensional Cartesian coordinate system. The RHS, deployed on the y-z plane, is centered at $[0, 0, 10]$ m with $10 \times 10$ elements. Faulty RHS elements are randomly distributed, ranging from $F=10$ to $F=60$. The number of feeds is $N_{RF}=6$. The BS serves $K=2$ UEs and one target. UEs are uniformly distributed in a circular region centered at $[40, 0, 1]$ m with radius 40 m, while the target is located at $[45, 45, 1.5]$ m, yielding an AoD ${\boldsymbol{\phi}} = {\left[ {{\phi _e},{\phi _a}} \right]^{\rm{T}}} = {\left[{-7.6^ \circ},{45^ \circ} \right]^{\rm{T}}}$. The system operates at 28 GHz with $P_{\max } = 33$ dBm. BS-UE channels follow Rician fading with K-factor 1. Path loss is modeled as $PL(d)=K_0(d/d_0)^{-\mu}$, where ${K_0} = (\lambda/4\pi)^2 \approx -60$ dB, $d_0 = 1$ m, ${\mu} = 2.2$, and $d$ is the link distance. The RHS element spacing is $\lambda /4$ with the refractive index $\kappa = \sqrt 3$. The SINR threshold and noise power are ${\gamma _{th}} = 10$ dB and $-80$ dBm, respectively.

\begin{figure*}[!t]
    \begin{minipage}[t]{0.33\linewidth}
    \centering
    \includegraphics[width=0.95\textwidth]{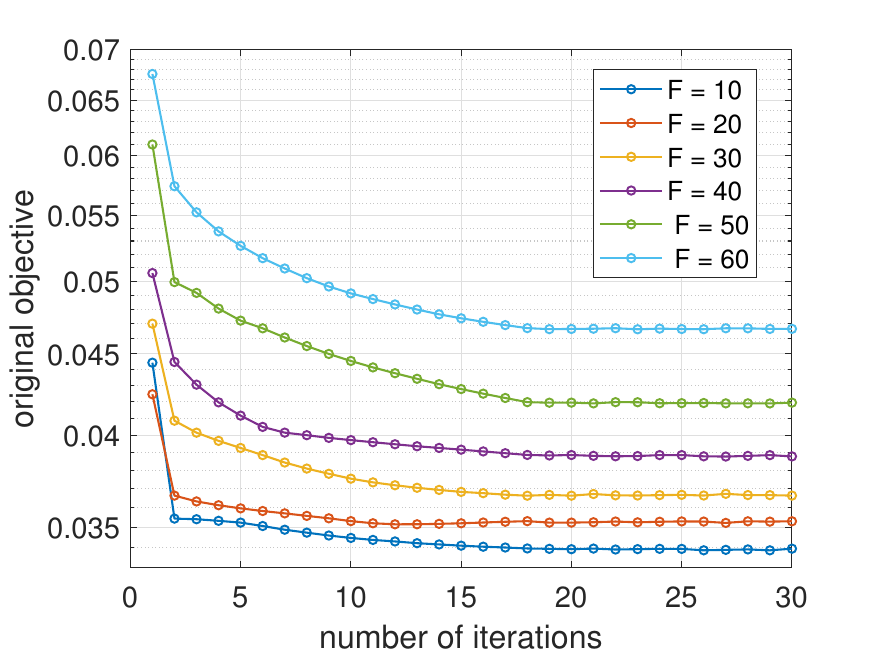}
    \caption{\small{Convergence performance of the proposed algorithm.}}
    \label{Re_conver}
    \end{minipage}
    \begin{minipage}[t]{0.33\linewidth}
    \centering
    \includegraphics[width=0.95\textwidth]{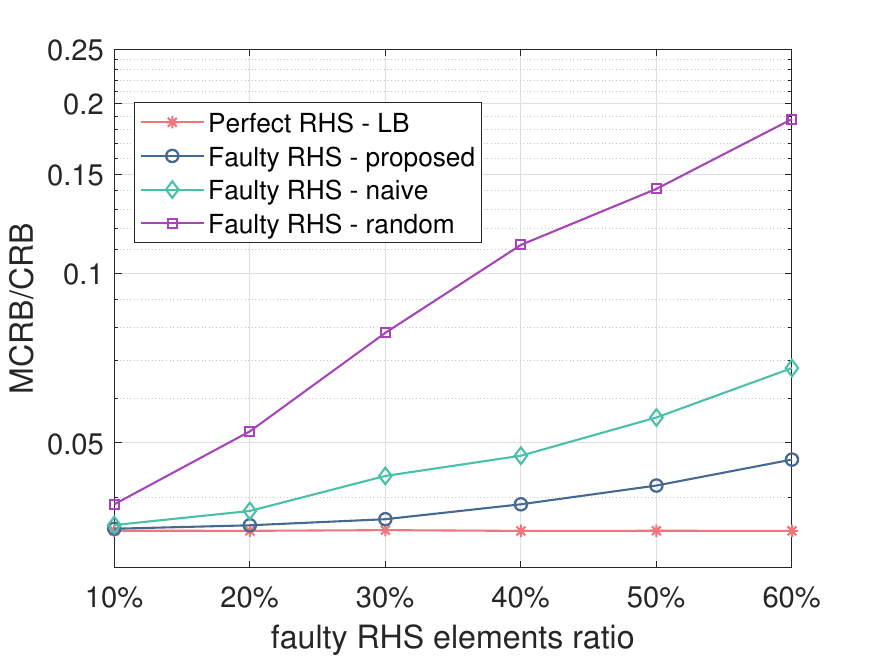}
    \caption{\small{MCRB/CRB under varying ratios of faulty elements.}}
    \label{Re_RF_MCRB}
    \end{minipage}
    \begin{minipage}[t]{0.33\linewidth}
    \centering
    \includegraphics[width=0.95\textwidth]{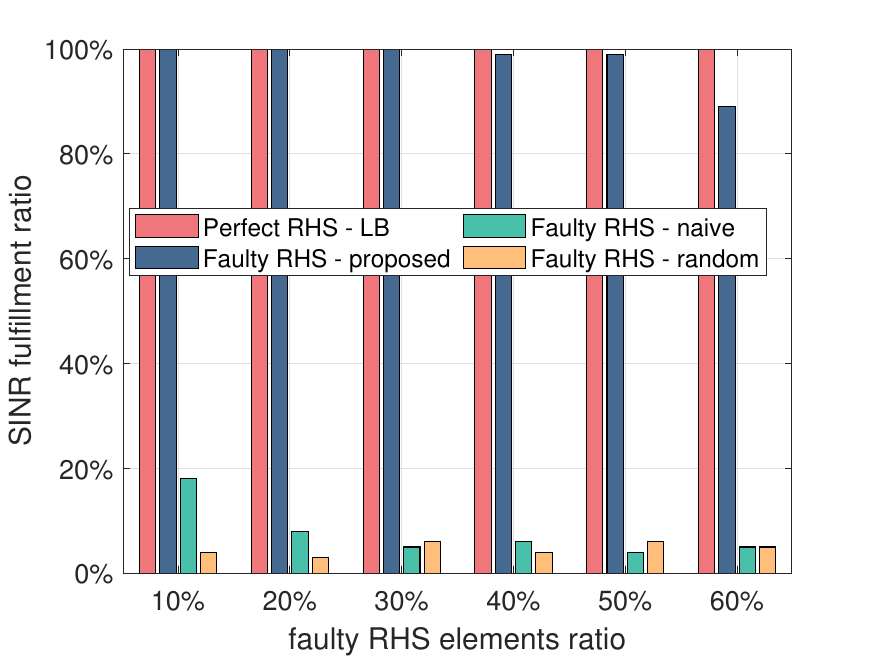}
    \caption{\small{SINR fulfillment ratio under varying ratios of faulty elements.}}
    \label{Re_RF_SINR}
    \end{minipage}
    \vspace{-0.6cm}
\end{figure*}
%%%%%%%%%%%%%%%%%%%% Convergence %%%%%%%%%%%%%%%%%%%%%%%
Fig.~\ref{Re_conver} illustrates the change in the original objectives, namely ${\rm{tr}}({{\rm{MCRB}}_\phi })$, under different numbers of faulty RHS elements ranging from 10 to 60, confirming the convergence of the proposed algorithm. 

Fig.~\ref{Re_RF_MCRB} and Fig.~\ref{Re_RF_SINR} present the sensing MCRB and communication SINR fulfillment ratios under varying ratios of faulty RHS elements. As shown in Fig.~\ref{Re_RF_MCRB}, when the ratio of faulty
\newpage
\noindent elements increases, the MCRB increases accordingly, meaning that the sensing performance degrades as faults accumulate. Compared to the \textit{naive} and \textit{random} schemes, the \textit{proposed} scheme reduces the MCRB loss by 13.7\% and 44.12\% on average, respectively, towards the performance bound \textit{LB}. This proves the effectiveness of optimizing the functional RHS elements. Moreover, in Fig.~\ref{Re_RF_SINR}, the \textit{LB} scheme always satisfies the SINR thresholds, which is expected due to the absence of element failure. The \textit{proposed} scheme can achieve mostly 100\% SINR fulfillment ratios, unless faulty RHS elements exceed 60\% of total elements. This degradation is reasonable since with more than 60\% of RHS elements faulty, the remaining 40\% functional elements are insufficient to effectively optimize system performance. In contrast, \textit{naive} and \textit{random} schemes fail to satisfy the SINR requirements.

%%%%%%%%%%%%%%%%%%%% Parameter %%%%%%%%%%%%%%%%%%%%%%%
% mean((CRBnaive-CRB)./CRBnaive) - mean((MCRB-CRB)./MCRB) = 0.1370
% mean((MCRBrndRIS-CRB)./MCRBrndRIS) - mean((MCRB-CRB)./MCRB) = 0.4412
% mean((MCRB-CRB)./MCRB)
% mean((CRBnaive-CRB)./CRBnaive)
% mean((MCRBrndRIS-CRB)./MCRBrndRIS)
\section{Conclusion}

This paper investigated a practical issue in the RHS-aided ISAC system, where a subset of the RHS elements experience failures with random amplitudes. We first derived the MCRB for sensing and the SINR for communication to quantify the performance degradation due to faulty RHS elements. To mitigate this degradation, we formulated an optimization problem to guarantee sensing and communication performances by jointly designing digital beamforming and holographic beamforming. Simulation results confirmed that the proposed scheme achieves an average MCRB improvement of 13.7\% over the fault-unaware baseline.

%\appendix
%\input{9Appendix}
\bibliographystyle{IEEEtran}
% Generated by IEEEtran.bst, version: 1.14 (2015/08/26)

%\IEEEtriggeratref{7} % Ref[7] 强制换栏
%\bibliography{biblioISAC}

\end{document}